\newcommand{\DTN} {NiCl$_2$$\cdot$4SC(NH$_2$)$_2$}
\newcommand{\DTNX} {Ni(Cl$_{1-x}$Br$_x$)$_2$$\cdot$4SC(NH$_2$)$_2$}

\newcommand{\aver}[1]{\left\langle #1 \right\rangle}

\documentclass[aps,pra,reprint,showpacs,longbibliography,superscriptaddress, citeonline]{revtex4-1}
\usepackage{amsmath,amssymb,bm}
\usepackage{graphicx}
\usepackage{xspace}
\usepackage{latexsym}
\usepackage{color}
\usepackage{hyperref}
\usepackage{placeins}

\begin{document}
\title{Spin waves near the edge of halogen substitution induced magnetic order in Ni(Cl$_{1-x}$Br$_x$)$_2$$\cdot$4SC(NH$_2$)$_2$}

\author{A.~Mannig}
    \affiliation{Laboratory for Solid State Physics, ETH Z\"{u}rich, 8093 Z\"{u}rich, Switzerland}
    \homepage{http://www.neutron.ethz.ch/}

\author{K.~Yu.~Povarov}
    \email{povarovk@phys.ethz.ch}
        \affiliation{Laboratory for Solid State Physics, ETH Z\"{u}rich, 8093 Z\"{u}rich, Switzerland}

\author{J.~Ollivier}
    \affiliation{Institut Laue-Langevin, 6 rue Jules Horowitz, 38042 Grenoble, France}

\author{A.~Zheludev}
    \affiliation{Laboratory for Solid State Physics, ETH Z\"{u}rich, 8093 Z\"{u}rich, Switzerland}

\begin{abstract}
We report an inelastic neutron scattering study of magnetic
excitations in a quantum paramagnet driven into a magnetically
ordered state by chemical substitution, namely
Ni(Cl$_{1-x}$Br$_x$)$_2$$\cdot$4SC(NH$_2$)$_2$ with $x=0.21(2)$. The
measured spectrum is well accounted for by the generalized spin wave
theory (GSWT) approach [M. Matsumoto and M. Koga,
\href{http://journals.jps.jp/doi/abs/10.1143/JPSJ.76.073709}{J.
Phys. Soc. Jap. \textbf{76}, 073709 (2007)}]. This analysis allows
us to determine the effective Hamiltonian parameters for a direct
comparison with those in the previously studied parent compound and
``underdoped'' system. The issue of magnon lifetimes due to
structural disorder is also addressed.

\end{abstract}

\date{\today}
\maketitle

\section{Introduction}

Long range magnetic order can in some cases be induced in disordered
quantum spin systems by a continuous tuning of exchange
constants~\cite{Sachdev_NPhys_2008_QCPs}. In such quantum phase
transitions (QPTs) a
  quantum
paramagnet transforms to a semiclassical N\'{e}el state via a
softening of the spin gap. The best known examples in real materials
are driven by applying hydrostatic pressure. Pressure induced
ordering has been extensively studied the the dimer systems
TlCuCl$_3$~\cite{OosawaFujisawa_JPSJ_2003_TlCuCl3PindDiff,Ruegg_PRL_2008_PindTlCuCl,Merchant_NatPhys_2014_PindTlCuCl}
and
PHCC~\cite{Thede_PRL_2014_uSRPHCC,PerrenMoeller_PRB_2015_PHCCpressurized,MannigMoeller_PRB_2016_PHCXmuons},
as well as in the single ion singlet compound
CsFeCl$_3$~\cite{KuritaTanaka_PRB_2016_CsFeCl3pressure,Hayashida_PRB_2018_CsFeClinelasticorder}.
These transitions differ from the better known magnetic field
induced ``Bose--Einstein condensation of
magnons''~\cite{Giamarchi_NatPhys_2008_BECreview,Zapf_RMP_2014_BECreview}
in the same
species~\cite{NikuniOshikawa_PRL_2000_TlCuCl3BEC,TodaFujii_PRB_2005_CsFeClfieldinduced,StoneBroholm_NJP_2007_PHCC}.
The crucial distinction is that the excitation spectrum remains
parabolic at BEC, and thus the dynamical exponent $z=2$. In
contrast, at the pressure-induced QPT the spectrum is linear,
implying $z=1$.

Pressure is not the only potential ``handle'' on the exchange
constants. Chemical substitution on non-magnetic sites presents an
alternative. It directly affects exchange and anisotropy parameters
or produces an effect of ``chemical
pressure''~\cite{YuYin_Nat_2012_DTNboseglass,*YuMiclea_PRB_2012_DTNXexponents,WulfHuvonen_PRB_2013_DTNXdiffraction,Povarov_PRB_2015_DTNXIN5}.
Recently, we have demonstrated that chemical modification can indeed
drive a $z=1$ QPT, namely  in the  anisotropic $S=1$ quantum
paramagnet \DTN\ known as
DTN~\cite{PovarovMannig_PRB_2017_DTNXcrit}. Upon Br substitution for
Cl in \DTNX\ [this modification is abbreviated as DTNX], the spin
gap decreases. At around $x_{c}\simeq0.15$ the spin-singlet ground
state is replaced by spontaneous N\'{e}el order. In particular, at
$x=0.21(2)>x_c$, the material undergoes long-range ordering at
$T_{N}=0.64$~K, albeit with a much reduced ordered moment
$\aver{S}\simeq0.3\mu_{B}$ at low temperatures
\cite{PovarovMannig_PRB_2017_DTNXcrit}. In the present work we
continue the investigation of this weakly ordered system, focusing
on spin excitations. We find that the measured spectrum is
remarkably well described by the so-called generalized spin wave
theory
(GSWT)~\cite{Matsumoto_JPSJ_2007_DTNHiggs,ZhangWierschem_PRB_2013_DTNdispersion,MunizKato_PTEP_2014_GSWTspecial}.
This allows us to determine the Hamiltonian parameters and discuss
the placement of this compound of the theoretical phase diagram. The
effect of disorder on the magnetic excitations is found to be minor.

\section{Material and experiment}

\subsection{DTNX: a short introduction}

\begin{figure}
  \includegraphics[width=0.5\textwidth]{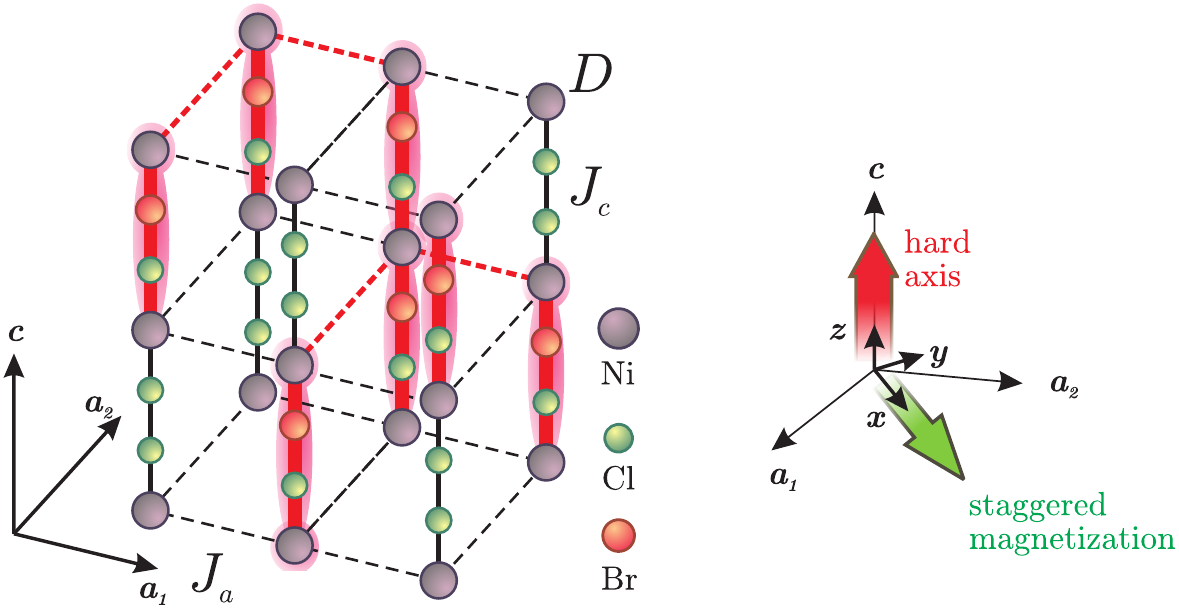}\\
  \caption{Left: a sketch of DTNX structure with only one tetragonal sublattice of the relevant ions shown. The principal Hamiltonian~(\ref{EQ:hamiltonian})
  is also illustrated: $D$ is the on-site planar anisotropy, solid bonds indicate
  stronger exchange along the ``chains'' $J_{c}$ and dashed bonds indicate weaker in-plane exchange $J_{a}$.
  Near the Br-substituted sites these values are affected. Right: the geometry conventions used in the paper. }\label{FIG:Structure}
\end{figure}

The physics of the parent compound \DTN\ is rather well
understood~\cite{Zapf_PRL_2006_BECinDTN,Zvyagin_PRL_2007_ESRinDTN,*YinXia_PRL_2008_DTNcritical,*BlinderDupont_PRB_2017_DTNNMR,WulfHuvonen_PRB_2015_DTNIntrinsicBroadening}.
The $S=1$ ions of Ni$^{2+}$ are bridged by two Cl ions into linear
chains, which run along the high-symmetry axis of the tetragonal
structure ($a=9.56$~\AA\ and $c=8.98$~\AA, see
Fig.~\ref{FIG:Structure}). Due to the body-centered $I4$ space group
there are two such tetragonal ``sublattices'', effectively decoupled
from each other magnetically.

A model magnetic Hamiltonian can be written as:

\begin{equation}\label{EQ:hamiltonian}
    \mathcal{\hat{H}}=\sum\limits_{\mathbf{r}}^{\text{Site}}D(\hat{S}^{z}_{\mathbf{r}})^{2}+
    \sum\limits_{\mathbf{r}}^{\text{Chain}}J_{c}(\hat{\mathbf{S}}_{\mathbf{r}}\cdot\hat{\mathbf{S}}_{\mathbf{r}+\mathbf{c}})+
    \sum\limits_{\mathbf{r},\mathbf{a_{1,2}}}^{\text{Plane}}J_{a}(\hat{\mathbf{S}}_{\mathbf{r}}\cdot\hat{\mathbf{S}}_{\mathbf{r}+\mathbf{a_{1,2}}})
\end{equation}

Here $\mathbf{a}_{1,2}$ and $\mathbf{c}$ are lattice translations as
in Fig.~\ref{FIG:Structure}. The vector $\mathbf{r}$ runs through
all Ni$^{2+}$ sites. The strongest contribution
to~Hamiltonian~(\ref{EQ:hamiltonian}) is the easy-plane single ion
anisotropy $D=0.7$~meV. Intrachain Heisenberg exchange is
$J_{c}=0.15$~meV is significantly stronger than inter-chain
interactions $J_{a}\simeq0.1J_{c}$. The planar anisotropy term
dominates and drives the system into a trivial  quantum disordered
state with $S^{z}=0$ for each ion.

There are two inequivalent halogen sites in the structure. In \DTNX\
the  Br substitute tends to occupy a particular one of
them~\cite{YuYin_Nat_2012_DTNboseglass,OrlovaBlinder_PRL_2017_DTNXNMR}.
This distorts the environment of Ni$^{2+}$ and strongly affects the
covalency of the Ni-halogen bonds. In turn, interactions are
strongly modified, especially $J_c$ for which the two-halide bond is
the primary superexchange mediator~\cite{Povarov_PRB_2015_DTNXIN5}.
Recent NMR~\cite{OrlovaBlinder_PRL_2017_DTNXNMR} and
theoretical~\cite{DupontCapponi_PRL_2017_DTNX2BEC,*Dupont_PRB_2017_BrDTNmicro}
studies suggest that these modification of the
Hamiltonian~(\ref{EQ:hamiltonian}) parameters are very local.

\subsection{Experimental details}

The bulk of the work reported here is inelastic neutron scattering
experiments on 99\% deuterated single crystal samples of \DTNX.
These were grown from aqueous solution using the temperature
gradient method as described
in~\cite{Yankova_PhilMag_2012_ReviewXtals,Wulf_2015_PhDthesis}. The
Br concentration in as-grown crystals was verified by means of
single-crystals x-ray diffraction on an APEX-II Bruker
diffractometer and determined to be  $x=0.21(2)$.

Time-of-flight inelastic neutron measurements were performed on the
IN5 spectrometer at Institute
Laue--Langevin~\cite{OllivierMutka_JPSJ_2011_IN5}. A sample
consisting of two co-aligned crystals with total mass of about
$0.2$~g was installed onto the cold finger of a $^3$He-$^4$He
dilution refrigerator.  The scattering plane was $(1,-1,0)$ as
defined by its normal. This provided access to the momentum
transfers of
$\mathbf{Q}=(h,~h,~l)=h(\mathbf{a}_{1}^{\ast}+\mathbf{a}_{2}^{\ast})+l\mathbf{c}^{\ast}$
type. The measurements were performed at a base temperature of about
$70$~mK using  neutrons with incident energies
$E_{\text{i}}=2.26$~meV. Scattering data were recorded with a
position-sensitive detector array for a sequence of 140 frames with
$1^{\circ}$ sample rotation steps.  All data were analyzed using
\textsc{Horace} software~\cite{Ewings_NucInstr_2016_HORACE}.

\section{Results and data analysis}

\subsection{Overview of the excitation spectrum}

\begin{figure*}
  \includegraphics[width=\textwidth]{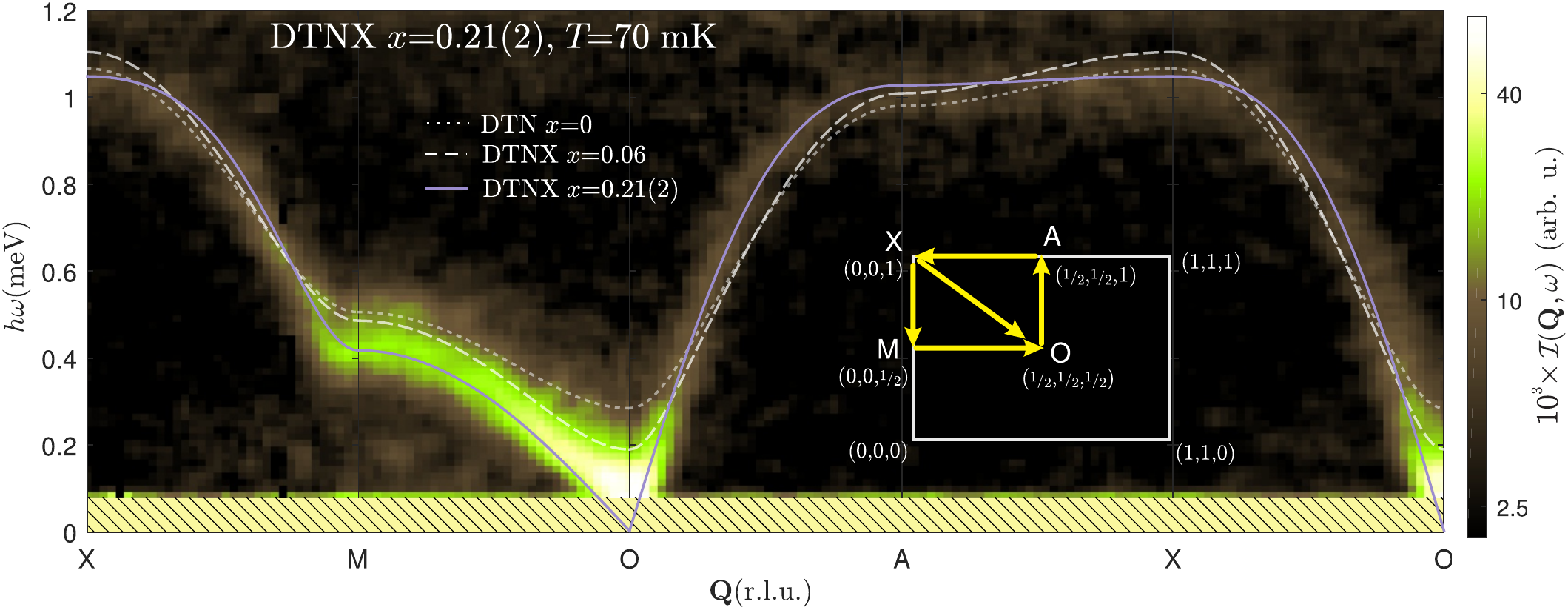}\\
  \caption{False color map of neutron scattering intensity measured in the $x=0.21(2)$ DTNX sample at $T\simeq70$~mK.
  The momentum transfer $\mathbf{Q}$ follow a specific trajectory between high symmetry points of
  the Brillouin zone (inset). The reference dispersion curves for $x=0$ and $x=0.06$
  materials~\cite{Zapf_PRL_2006_BECinDTN,Povarov_PRB_2015_DTNXIN5} are shown as dotted and dashed lines.
  The solid line is a dispersion fit to the present data, as described in the text. Hatching marks the region dominated by parasitic
  quasielastic scattering. }\label{FIG:bigpanelplot}
\end{figure*}

An overview of the measured excitation spectrum is given in
Fig.~\ref{FIG:bigpanelplot}. It shows a false color plot of neutron
scattering intensities as a function of momentum and energy
transfer. As indicated in the inset, the momentum transfer follows a
sequence of high symmetry directions in the $(h,~h,~l)$ plane. The
spectrum is clearly dominated by a single excitation branch, which
remains underdamped in the entire zone. Overall, its dispersion is
not dissimilar to that previously measured in the parent compound
$x=0$~\cite{Zapf_PRL_2006_BECinDTN} and for
$x=0.06$~\cite{Povarov_PRB_2015_DTNXIN5} (dotted and dashed lines in
Fig.~\ref{FIG:bigpanelplot}). The key difference is that in the
present compound there is no excitation gap. This is consistent with
notably different thermodynamics, as compared to $x<x_c$
(``underdoped'') materials~\cite{PovarovMannig_PRB_2017_DTNXcrit}.

\subsection{Theoretical approach}

As mentioned, the ordered moment in the present $x=0.21(2)$ material
is significantly reduced compared to the classical expectation  of
$2\mu_B$ per Ni$^{2+}$. Under these circumstances, the standard spin
wave theory reaches its applicability limit. To quantitatively
analyze the data we instead employed the so-called GSWT
approach~\cite{ZhangWierschem_PRB_2013_DTNdispersion,MunizKato_PTEP_2014_GSWTspecial}.
We follow the particular formulation for a DTN-like
Hamiltonian~(\ref{EQ:hamiltonian}) developed by Matsumoto and
Koga~\cite{Matsumoto_JPSJ_2007_DTNHiggs}. This method utilizes a
basis of local states and is similar to ``bond operator
theory''~\cite{SachdevBhatt_PRB_1990_LagrangeDimers,MatsumotoNormand_PRB_2004_TlCuClz1z2}
used for treating dimerized spin systems. In the quantum
paramagnetic phase the GSWT spectrum features two degenerate bosonic
modes with dispersion relations identical to those obtained in the
random phase approximation
(RPA)~\cite{LindgardSchmid_PRB_1993_strongD_RPA}. The ordered state
is then treated as the quantum mechanical condensate of such
excitations. Here there are three new types of quasiparticles: two
transverse spin wave modes ($yy$) and ($zz$) (see schematic in
Fig.~\ref{FIG:Structure}), and one amplitude mode ($xx$).
 The
transverse modes are gapless. Their dispersions are related by a
translation by the magnetic propagation vector
$\mathbf{Q}_{0}=(1/2,~1/2,~1/2)$:
$\hbar\omega_{yy}(\mathbf{Q})=\hbar\omega_{zz}(\mathbf{Q}+\mathbf{Q}_{0})$.
The equal-time structure factor (intensity)
$\mathcal{S}_{yy}(\mathbf{Q})$ is generally much larger than
$\mathcal{S}_{zz}(\mathbf{Q})$, since the latter corresponds to spin
fluctuations along the hard axis. The longitudinal mode has a gap,
which is proportional to the ordered moment. Its intensity  tends to
be smaller than that for the $yy$ spin wave. The exact GSWT
expressions for the corresponding  dispersion relations and
structure factors are given in Appendix A.

\subsection{Data analysis}

\begin{figure}
    \includegraphics[width=0.5\textwidth]{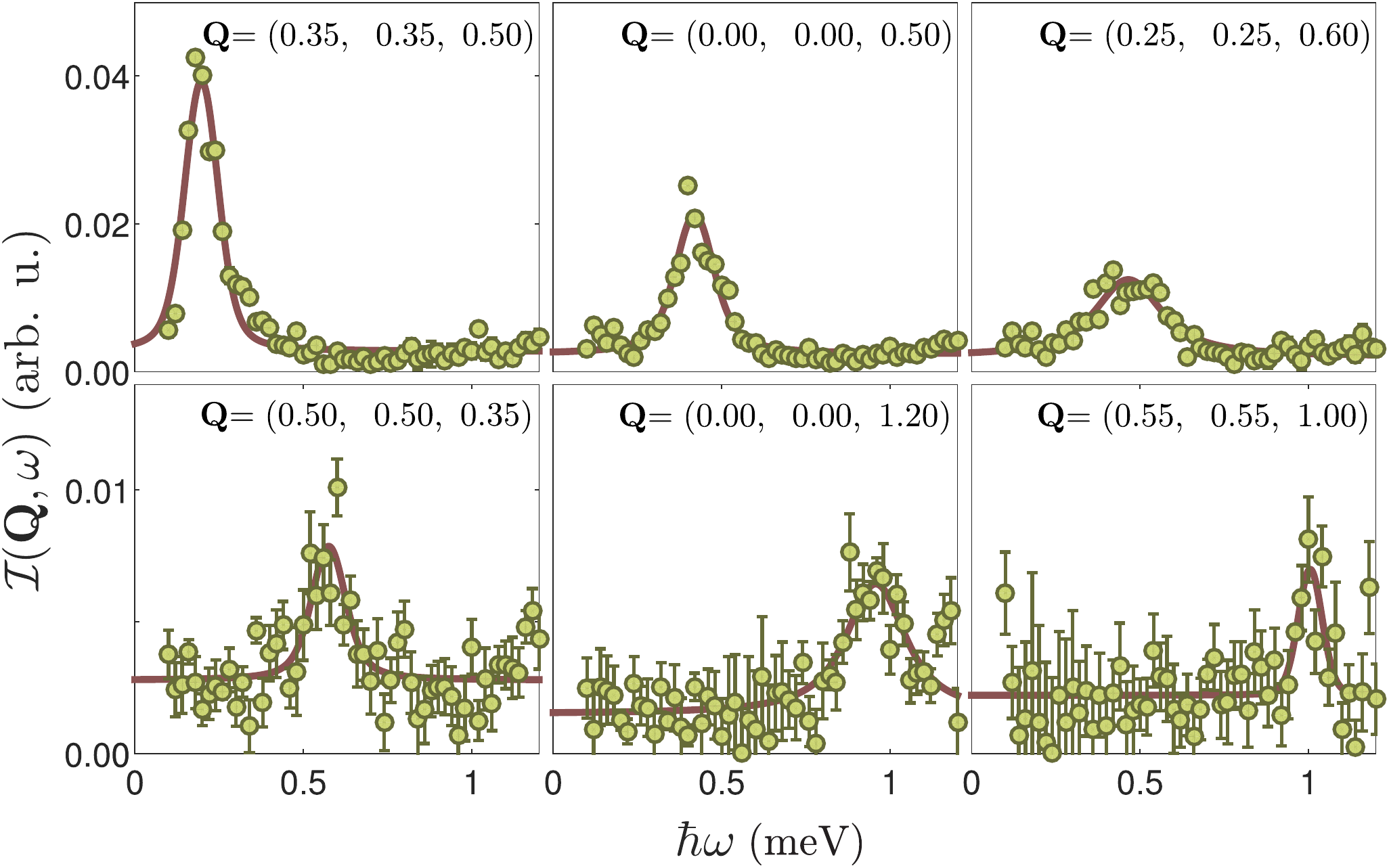}\\
    \caption{Representative constant-$\mathbf{Q}$ cuts of the
    measured neutron intensity (symbols). The data is truncated
    below $\hbar\omega=0.1$ and above $\hbar\omega=1.2$~meV, where instrument background
    dominates. Solid lines are fits to individual cuts, as described in the text.}\label{FIG:Selectedcuts}
\end{figure}

\begin{figure}
    \includegraphics[width=0.5\textwidth]{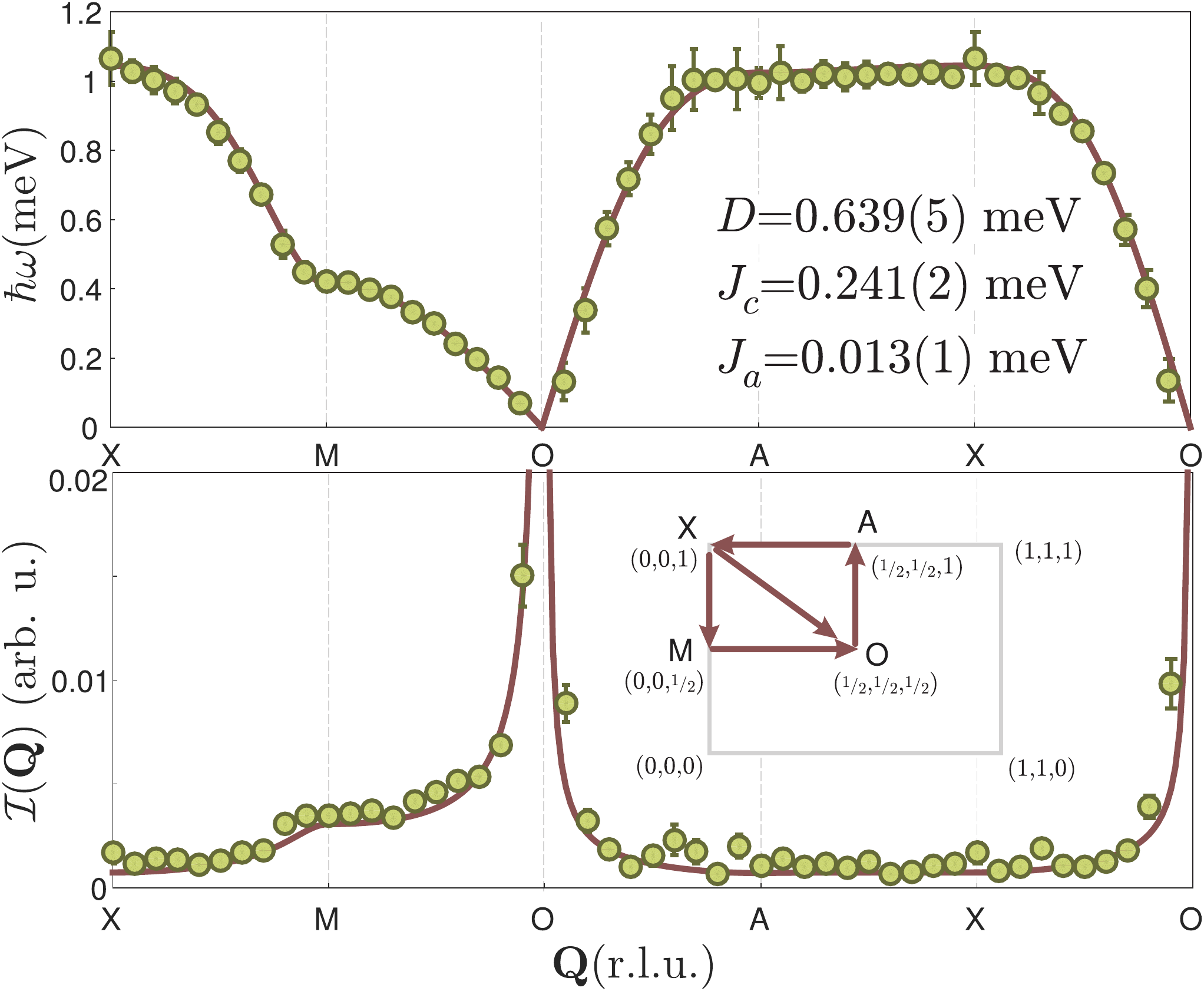}\\
    \caption{Symbols: dispersion (top) and intensities $\mathcal{I}(\mathbf{Q})=|F(Q)|^{2}P_{yy}(\mathbf{Q})\mathcal{S}_{yy}(\mathbf{Q})$
    (bottom) of magnetic excitations determined in fits to individual constant-$\mathbf{Q}$ cuts.
    In the dispersion plot, the error bars are actually the intrinsic line widths determined
        from the same fits. The solid line is the GSWT calculation based on Hamiltonian parameters
        determined in a global fit to the measured dispersion relation.}\label{FIG:pointpanels}
\end{figure}

In order to quantify the dispersion, intrinsic width and intensity
of the observed excitations, we analyzed individual
constant-$\mathbf{Q}$ cuts of the data for a grid of wave vectors
with a step of $\delta Q=0.05$~r.l.u. in both $h$ and $l$
directions. At each wave vector the fitting function was a Voigt
profile. Its Gaussian component was the calculated energy-dependent
energy resolution of the
spectrometer~\cite{Lowde_JNE_1960_TOFresolution,*Lechner_1985_TOFresolution}.
The Lorentzian component represented the intrinsic excitation width
and was one of the fit parameters. Also fitted was the peak
position, an intensity prefactor and a a flat background.
Representative examples of such fits for a few representative points
are given in Fig.~\ref{FIG:Selectedcuts}. For high-symmetry
reciprocal space directions the fitted peak positions and
intensities are plotted against wave vector transfer in
Fig.~\ref{FIG:pointpanels}.

To obtain the Hamiltonian parameters the dispersion relation
determined on the entire wave vector grid was fit using the GSWT
result. We attribute the observed scattering to the $yy$ excitation
branch. The best fit is obtained with $D=0.639(5)$~meV,
$J_{c}=0.241(2)$~meV, and $J_{a}=0.013(1)$~meV. The dispersion
relation along high symmetry directions calculated with these values
is plotted in a solid line in Fig.~\ref{FIG:pointpanels} (top panel)
to illustrate the excellent level of agreement.

A GSWT calculation with these parameters can reproduce the measured
intensities as well. This defines the neutron polarization factor,
as neutrons are only scattered by magnetization components that are
transverse to the momentum transfer. Since a macroscopic sample is
bound to split into domains with different orientations of the
ordered moment in the tetragonal plane, this effect needs to be
averaged accordingly. In our analysis we thus multiplied the mode
intensities calculated with GSWT by the polarization factor
$P_{yy}(\mathbf{Q})=1-\left(2\pi ha/Q\right)^{2}$. An additional
correction was the Ni$^{2+}$ magnetic form factor $|F(Q)|^{2}$ that
we included in our calculation within the dipole
approximation~\cite{Prince_2004_XtalTables}. With just one
additional fit parameter, namely a single overall scale factor, the
GSWT model with exchange and anisotropy constants as obtained in
dispersion fit gives an excellent agreement with the measurement
(solid line in the bottom panel of Fig.~\ref{FIG:pointpanels}).

\begin{figure}
  \includegraphics[width=0.4\textwidth]{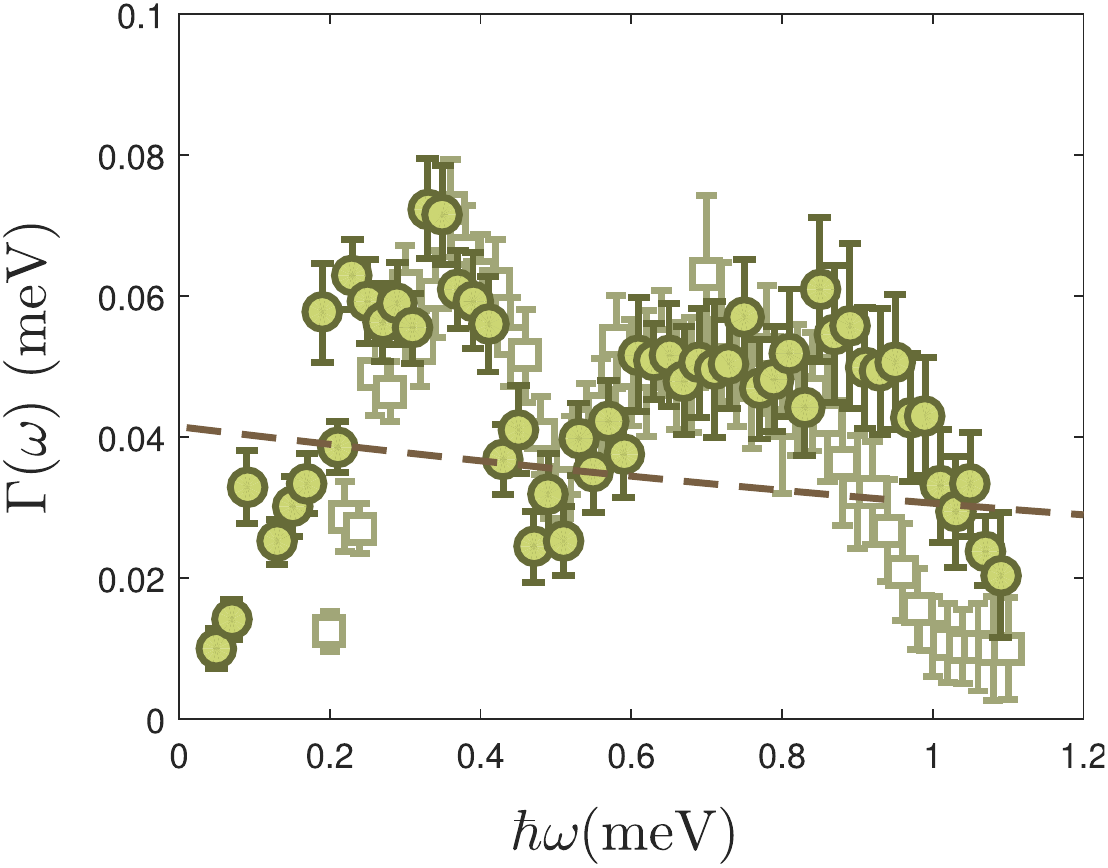}\\
  \caption{Measured distribution of the intrinsic linewidth as the function of
  excitation energy. Filled circles are the present data for $x=0.21$ DTNX.
  Open squares are for $x=0.06$~\cite{Povarov_PRB_2015_DTNXIN5}. The dashed line is the calculated spectrometer
  resolution in both cases.}\label{FIG:damping}
\end{figure}

Our data analysis also yields the intrinsic linewidth
$\Gamma(\mathbf{Q})$ of excitations as a function of wave vector.
For a direct comparison with previous studies, we choose to plot the
energy dependence of the linewidth $\Gamma(\omega)$, averaged over
the Brillouin zone. For our present sample this quantity is shown in
Fig.~\ref{FIG:damping} in filled circles. Previously published data
for  the ``underdoped'' $x=0.06$ material are plotted in open
squares. For reference, the energy resolution of the spectrometer
(the same in both studies) is plotted in a dashed line.

\section{Discussion}
\subsection{Other GSWT modes}

\begin{figure*}
  \includegraphics[width=1\textwidth]{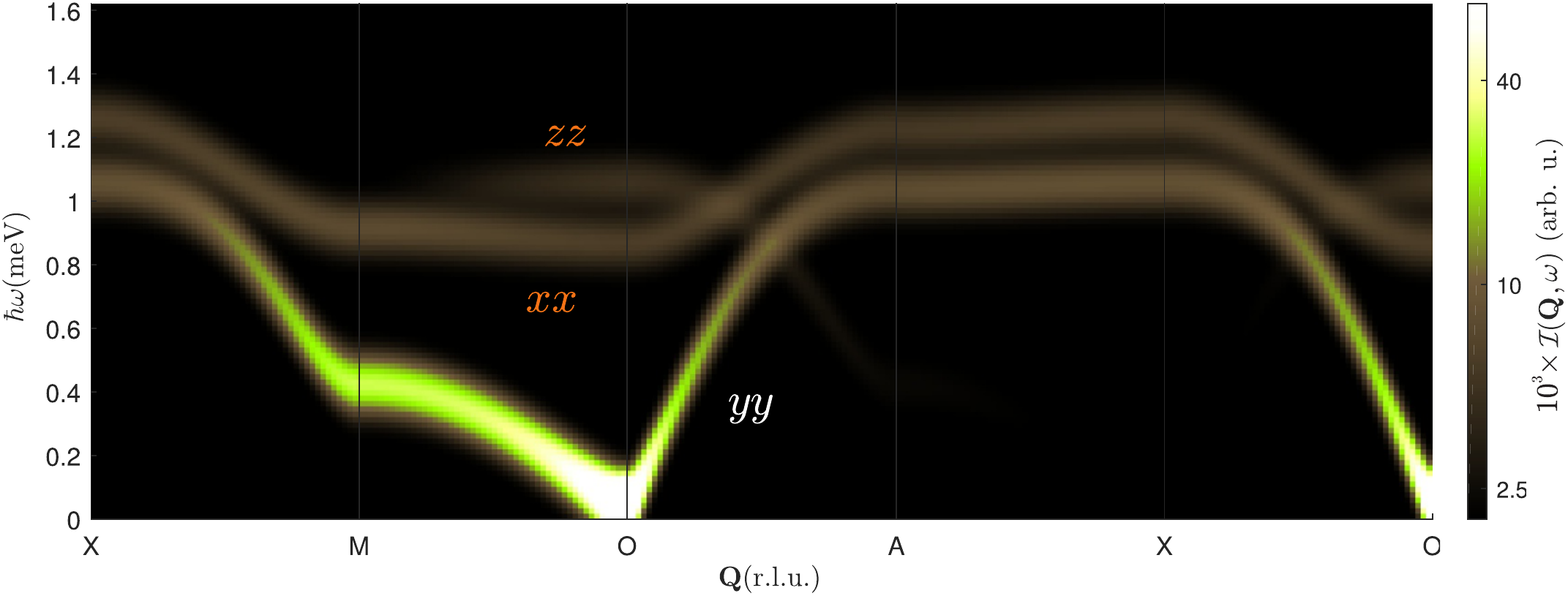}\\
  \caption{Simulated GSWT spectrum with the background $\mathcal{B}(\mathbf{Q})=2\cdot10^{-3}$, convoluted with $\sigma=50$~$\mu$eV Gaussian.
  Polarization factors and magnetic ion form factor are also taken into account.}\label{FIG:SWsimulation}
\end{figure*}

The agreement of the GSWT predictions for the $yy$ mode with
experiment is remarkable, but what about the other two excitation
branches? In Fig.~\ref{FIG:SWsimulation} we show the GSWT
calculation for all three polarizations, based on the Hamiltonian
parameters obtained in the fit above. The polarization factors for
each mode and the magnetic form factor are accounted for as
appropriate. All modes in this simulation have zero intrinsic line
widths, as appropriate for GSWT. This simulation tells us that the
out of plane transverse $zz$ mode could well be too weak to be
observable in the present experiment. However, the longitudinal $xx$
mode would be strong enough to be seen, if only it were underdamped.
It is well understood however that a sharp longitudinal excitation
is an artifact of GSWT. In fact, longitudinal modes in
antiferromagnets with gapless spin waves  are known to be prone to
decays into transverse
modes~\cite{PodolskyAuerbach_PRB_2011_HiggsVisibility,ZhitomirskyChernyshev_RMP_2013_DecayReview}.
Being overdamped, they cannot be even associated with a peak feature
in the spectrum. In the few
materials~\cite{Merchant_NatPhys_2014_PindTlCuCl,HongMatsumoto_NPhys_2017_LadderHiggs,JainKrautlocher_NPhys_2017_2DhiggsCa2RuO4}
where longitudinal modes are observed, they are stabilized by
Ising-like anisotropy effects absent in DTNX. We conclude that
detecting only one of the excitation branches predicted by GSWT is
actually not that surprising.

\subsection{Hamiltonian parameters}

\begin{table}
\centering
    \begin{tabular}{c l c l c l c}
    \hline  \hline
    Hamiltonian~(\ref{EQ:hamiltonian}) & & clean & & ``underdoped''  & &
    ``overdoped''  \\
    parameters   & & $x=0$ & & $x=0.06$  & & $x=0.21(2)$ \\
     (meV) & & Ref.~~\cite{Zapf_PRL_2006_BECinDTN} & &
     Ref.~\cite{Povarov_PRB_2015_DTNXIN5} & & present work\\
        \hline
        $D$     & & 0.780(3)    & &  0.792(3) & & 0.639(5) \\
        $J_{c}$ & & 0.141(3)    & &  0.155(1) & & 0.241(2) \\
        $J_{a}$ & & 0.014(1)    & &  0.0158(3)& & 0.013(1) \\
        \hline \hline
    \end{tabular}
    \caption{Parameters of Hamiltonian~(\ref{EQ:hamiltonian}) for various members of DTNX family experimentally determined by means of
    GSWT analysis of inelastic neutron scattering data.}\label{TAB:params}
\end{table}

As already mentioned, in the quantum paramagnetic phase GSWT is
equivalent to the RPA. This allows a meaningful comparison of the
Hamiltonian parameters that we obtain for the ``overdoped''
$x=0.21(2)>x_c$ material to those previously determined for the
parent compound ~\cite{Zapf_PRL_2006_BECinDTN} and for the
``underdoped'' $x=0.06$ system~\cite{Povarov_PRB_2015_DTNXIN5}.

\begin{figure}
  \includegraphics[width=0.5\textwidth]{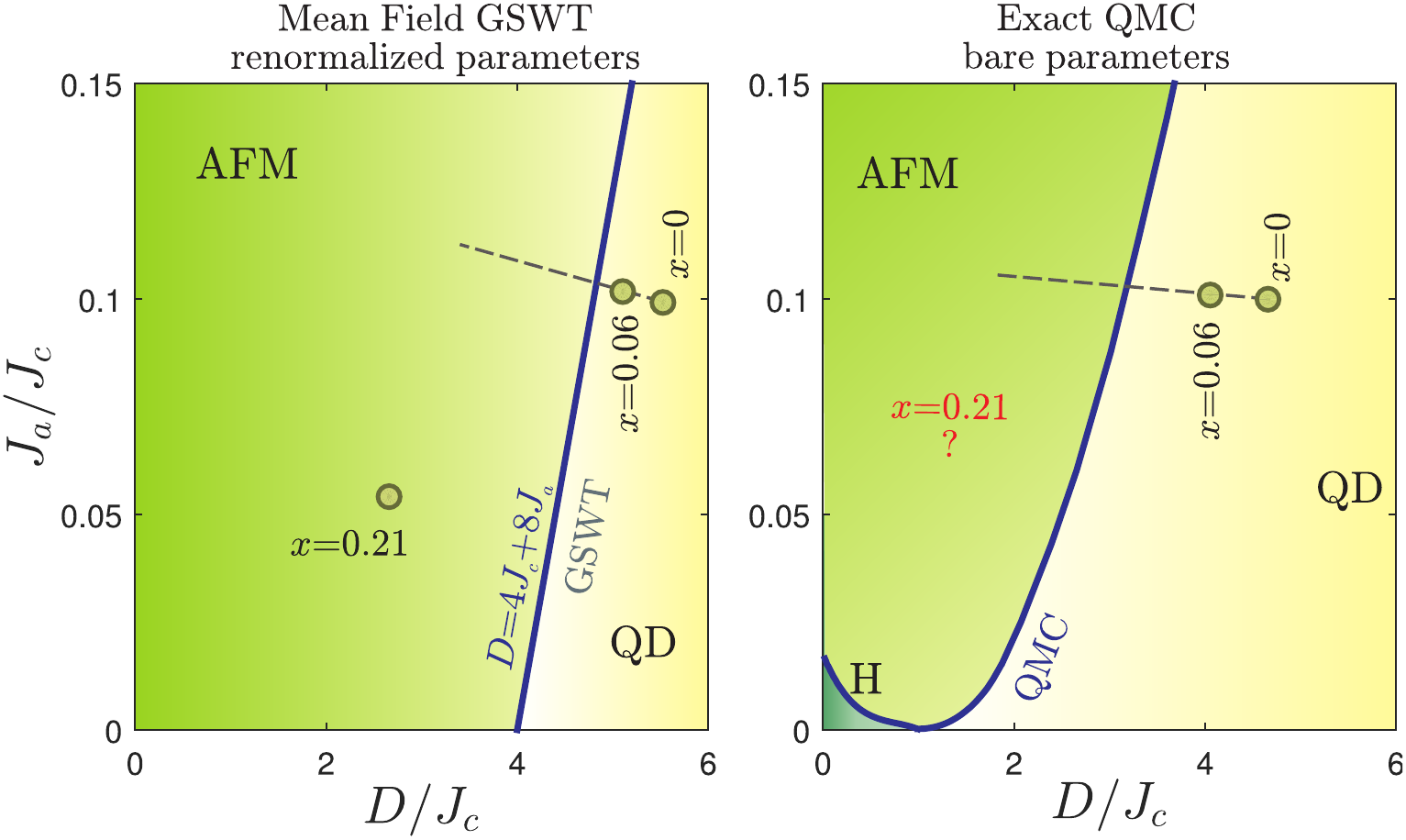}\\
  \caption{Ground state phase diagrams of Hamiltonian~(\ref{EQ:hamiltonian}). Left panel: Phase diagram in the mean field approximation
  including quantum disordered (QD) and antiferromagnetically ordered (AF) phases~\cite{Matsumoto_JPSJ_2007_DTNHiggs}. The points, locating three
  different DTNX family members
  on this phase diagram are obtained from GSWT \emph{without} taking the quantum corrections into account. Right panel: the exact phase diagram, obtained
  by Quantum Monte--Carlo (QMC) calculations. A new Haldane phase (H) appears in the decoupled Heisenberg chain limit. The experimental points for $x=0$ and $x=0.06$
  materials of DTNX family now correspond to the estimated bare microscopic
   parameters~\cite{Zapf_PRL_2006_BECinDTN,Povarov_PRB_2015_DTNXIN5}. Dashed line shows the linear low bromine concentration trend in both figures.}\label{FIG:phasediagrams}
\end{figure}

This comparison is made in Table~\ref{TAB:params}. The left panel of
Fig.~\ref{FIG:phasediagrams} positions the three compounds on the
phase diagram calculated with the Mean Field (MF) or GSWT
approximations. The MF boundary between the gapped and ordered
phases, i.e., the line of gap closure, corresponds to $D=4J_c+8J_a$.
In this context, the $x=0.21(2)$ material lands deep inside the
ordered phase. The complication is that the RPA and GSWT exchange
constants are known to be strongly renormalized compared to actual
values. For the gapped compounds this renormalization can be
accounted for by the self-consistent ``Lagrange multiplier
method''~\cite{SachdevBhatt_PRB_1990_LagrangeDimers,ZhangWierschem_PRB_2013_DTNdispersion}.
The phase space of actual, rather than renormalized, Hamiltonian
parameters is shown in the right panel of
Fig~\ref{FIG:phasediagrams}. This plot also shows the numerically
computed phase boundaries for weakly-coupled $S=1$ spin chains with
planar anisotropy
~\cite{SakaiTakahashi_PRB_1990_DTNlikeGS,WierschemSengupta_MPhysLettB_2014_DTNlikeGS,*WierschemSengupta_PRL_2014_DTNlikeGS}.
Unfortunately, this procedure cannot be applied to determine the
actual Hamiltonian parameters in our ``overdoped'' system, which is
already in the ordered
phase~\cite{ZhangWierschem_PRB_2013_DTNdispersion}.

\subsection{Disorder}

\DTNX\ obviously contains a large amount of structural disorder
which, as mentioned above, translates into a randomness of local
Hamiltonian parameters. While one could expect this randomness to be
stronger in our $x=0.21(2)$ material than in the previously studied
$x=0.06$ system, the comparison of intrinsic line widths
(Fig.~\ref{FIG:damping}) show that the two are comparable. Moreover,
despite rather distinct ground states, the line width distributions
are very similar, with maxima around $0.3$ and $0.8$~meV, and a
starpening of magnon peaks at the top and bottom of the spectrum. We
conclude that across the concentration range the magnon instability
due to randomness must be governed by a single mechanism.

\section{Summary}

The three take-home messages of this study are: 1) Transverse spin
excitations in the entire series of DTNX materials, both
``underdoped'' and ``overdoped'', are remarkably well described by
GSWT. 2) The underdamped longitudinal mode, a known artefact of
GSWT, is absent in the DTNX. 3) The effect of chemical disorder is
rather subtle and for the most part amounts to a moderate broadening
of magnons.

\acknowledgments

This work was supported by Swiss National Science Foundation,
Division II. We would like to thank Dr. S. Gvasaliya
(ETH~Z\"{u}rich) for assistance with the sample alignment for the
neutron experiment.

\appendix
\section{GSWT exact results}
\label{SEC:swappendix}

In this Appendix we recite the GSWT theoretical results for the
DTN-like material obtained by Matsumoto and
Koga~\cite{Matsumoto_JPSJ_2007_DTNHiggs}. We would like to start
with the excitations in the ordered phase. First, the following
auxiliary notation are introduced:

\begin{equation}\label{EQ:uv_definition}
    u,v=\sqrt{\frac{1}{2}\left(1\pm\frac{D}{4J_{c}+8J_{a}}\right)},
\end{equation}

\begin{equation}\label{EQ:gamma_definition}
   \gamma(\mathbf{Q})=2J_{c}\cos(\mathbf{Q\cdot c})+2J_{a}\left[\cos(\mathbf{Q\cdot a_{1}})+\cos(\mathbf{Q\cdot
   a_{2}})\right].
\end{equation}

Note, that within GSWT $2\gamma(0)=4J_{c}+8J_{a}$ corresponds to the
critical value of single-ion anisotropy $D_c$ at which the
antiferromagnetic order is suppressed.

Then, Eqs.~(\ref{EQ:uv_definition}) and (\ref{EQ:gamma_definition})
are plugged into the following four terms related to longitudinal
($L$) and transverse ($T$) excitation channels:

\begin{eqnarray}\label{EQ:epsdelta_definition}
    \nonumber
    \epsilon_{L}(\mathbf{Q})&=&(u^{2}-v^{2})D+4u^{2}v^{2}(4J_{c}+8J_{a})\\
    \nonumber
    &&+(u^{2}-v^{2})^{2}\gamma(\mathbf{Q}),\\
    \delta_{L}(\mathbf{Q})&=&(u^{2}-v^{2})^{2}\gamma(\mathbf{Q}),\\
    \nonumber
    \epsilon_{T}(\mathbf{Q})&=&u^{2}D+2u^{2}v^{2}(4J_{c}+8J_{a})+(u^{2}-v^{2})\gamma(\mathbf{Q}),\\
    \nonumber
    \delta_{T}(\mathbf{Q})&=&\gamma(\mathbf{Q}).
\end{eqnarray}

Finally, the terms~(\ref{EQ:epsdelta_definition}) are used to
construct the dispersion relations and the corresponding equal-time
structure factors (intensities) of the spin wave excitations:

\begin{equation}\label{EQ:swmodeL}
    \hbar\omega_{xx}(\mathbf{Q})=\sqrt{\epsilon_{L}(\mathbf{Q})^{2}-\delta_{L}(\mathbf{Q})^{2}},
\end{equation}

\begin{equation}\label{EQ:swmodeTy}
    \hbar\omega_{yy}(\mathbf{Q})=\sqrt{\epsilon_{T}(\mathbf{Q})^{2}-\delta_{T}(\mathbf{Q})^{2}},
\end{equation}

\begin{equation}\label{EQ:swmodeTz}
    \hbar\omega_{zz}(\mathbf{Q})=\sqrt{\epsilon_{T}(\mathbf{Q}+\mathbf{Q}_{0})^{2}-\delta_{T}(\mathbf{Q}+\mathbf{Q}_{0})^{2}},
\end{equation}

\begin{equation}\label{EQ:swintXX}
    \mathcal{S}_{xx}(\mathbf{Q})=A(u^{2}-v^{2})^{2}\sqrt{\frac{ \epsilon_{L}(\mathbf{Q})-\delta_{L}(\mathbf{Q})}{\epsilon_{L}(\mathbf{Q})+\delta_{L}(\mathbf{Q})}},
\end{equation}

\begin{equation}\label{EQ:swintYY}
    \mathcal{S}_{yy}(\mathbf{Q})=A u^{2}\sqrt{\frac{ \epsilon_{T}(\mathbf{Q})-\delta_{T}(\mathbf{Q})}{\epsilon_{T}(\mathbf{Q})+\delta_{T}(\mathbf{Q})}},
\end{equation}

\begin{equation}\label{EQ:swintZZ}
    \mathcal{S}_{zz}(\mathbf{Q})=A v^{2}\sqrt{\frac{
    \epsilon_{T}(\mathbf{Q+Q_{0}})+\delta_{T}(\mathbf{Q+Q_{0}})}{\epsilon_{T}(\mathbf{Q+Q_{0}})-\delta_{T}(\mathbf{Q+Q_{0}})}}.
\end{equation}

In the equations above $A$ is the overall normalization parameter,
used for direct comparison with the experimental neutron scattering
intensities.

In case of quantum paramagnetic phase the $zz$ excitation vanishes,
while $xx$ and $yy$ excitations converge to a single doubly
degenerate mode. Its dispersion is simply:

\begin{equation}\label{EQ:RPAmode}
    \hbar\omega_{PM}(\mathbf{Q})=\sqrt{D^{2}+2D\gamma(\mathbf{Q})}.
\end{equation}

The corresponding intensity is given by:

\begin{equation}\label{EQ:swintRPA}
    \mathcal{S}_{PM}(\mathbf{Q})=A \frac{D}{\hbar\omega_{PM}(\mathbf{Q})},
\end{equation}

with $A$ being the same as in Eqs.~(\ref{EQ:swintXX}) --
(\ref{EQ:swintZZ}).

\bibliography{d:/The_Library}
\end{document}